\documentstyle[preprint,aps]{revtex}

\begin{document}
\draft

\title{Magnetization of Disordered Ballistic Quantum Billards}
\author{ Daniel Braun$^1$, Yuval Gefen$^{1,2}$ and Gilles Montambaux$^1$}
\address{$^1$ Laboratoire de Physique des Solides, Associ\'e au CNRS,
Universit\'e  Paris--Sud, 91405 Orsay, France\\
$^2$ Department of Condensed Matter Physics, The Weizmann Institute of
Science, Rehovot 76100, Israel}
\date{May 1994}
\maketitle
\begin{abstract}
We study the magnetic response of mesoscopic quantum dots in the ballistic
regime where the mean free path $l_e$ is larger that the size $L$ of the
sample, yet smaller than $L (k_F L)^{d-1}$. In this regime, disorder
 plays an important role. Employing a semiclassical picture we calculate
the contribution of long
trajectories which are strongly affected by static disorder and  which
differ sharply
from those of clean systems.
In the case of a magnetic field,
they give rise to a large linear paramagnetic
susceptibility (which is disorder independent), whose magnitude is in
agreement with recent
experimental results. In the case of a Aharonov-Bohm flux, the susceptibility
is disorder dependent and is proportional to the mean free path as in the
diffusive regime. We also discuss the corresponding non--linear
susceptibilities.\\
\end{abstract}
\newpage

\section{Introduction}

Anticipating and following the first experiments  on persistent
currents in mesoscopic rings,  there has been a large number of
theoretical studies of the magnetization of phase coherent
disordered electronic systems in the diffusive regime, i.e.
when the elastic mean free path $l_e$ is smaller than the
characteristic linear size of the sample $L$. On the other hand, the most
recent experiments\cite{benoit,levy} correspond to a situation where
$l_e$ equals or exceeds $L$. The present analysis is concerned with a regime
which has been christened {\it
ballistic}. Originally, this name simply referred to the regime $l_e >
L$. This, however,  turned out to be somewhat misleading as one has to
separate the range where impurity scattering is significant
({\it ballistic} regime) from the {\it clean} or {\it perturbative} regime.
The latter, defined by $l_e>L(k_FL)^{d-1}$, is  as yet
experimentally inaccessible. Single particle energy levels in the
perturbative regime are closely
approximated by those of a perfectly clean system of the same
geometry. In other words, impurity mixing of the levels is small.

In the present work we consider the {\it ballistic} regime $L
< l_e < L(k_F L)^{d-1}$, addressing the important issue of the
average magnetization (or average persistent current in the
case of an
Aharonov-Bohm geometry) and the weak-field susceptibility
\cite{Sivan87,altland}.
Within the framework of non-interacting electrons, it has been
stressed that some care has to be given to the procedure of
averaging\cite{bouchiat}. More specifically, it
has been shown that, in the mesoscopic regime, there is an
important contribution to the magnetic response under
"canonical conditions", i.e. when the number of particles in
each sample is kept fixed as the magnetic flux is varied.
 This additional contribution reads

\begin{equation}
\langle M(H)\rangle=-\frac{\Delta}{2}\frac{\partial}{\partial H}
\langle\delta N^2(\mu,H)\rangle\,,
                                                       \label{M}
\end{equation}
where $\langle\delta N^2(\mu,H)\rangle$ is the typical field dependent
 sample to
sample fluctuation in the number of levels below the chemical
potential $\mu$\cite{schmid,oppen1,AGI,AGIM}. $\Delta$ is the mean level
spacing.

In the diffusive regime, only the disorder average has been considered.
 It turns
out that, in the ballistic regime, disorder averaging may not be the
 sole contribution, and that an additional term arises
which is associated with size averaging \cite{altland}.
 More precisely, the quantity to be
considered is:

\begin{equation}
\langle\delta N^2(\mu,H)\rangle =
\int_0^{\mu}\int_0^{\mu}K(\varepsilon, \varepsilon')
d\varepsilon d\varepsilon' \ ,
                                                               \label{dN}
\end{equation}
$K(\varepsilon, \varepsilon')$ being the field dependent
density of states (DOS) autocorrelation function:

\begin{equation}
K(\varepsilon, \varepsilon') = \langle \rho(\varepsilon)
\rho(\varepsilon') \rangle_{d,L} - \langle \rho(\varepsilon) \rangle_{d,L}
 \langle \rho(\varepsilon') \rangle_{d,L} \ .
                                                                    \label{K}
\end{equation}
The notation $\langle ...
\rangle_{d,L}$ indicates averaging over both disorder and system
size\footnote{To discard any structure in the DOS, the system
size has to be averaged over a range $\delta L \gtrsim \lambda_F$,
where $\lambda_F$ is the Fermi wave length}.
This correlation function may be rewritten as:
\begin{equation}
K(\varepsilon, \varepsilon') =
\Big\langle K_d(\varepsilon -
\varepsilon')\Big\rangle_L+K_L^{\prime}(\varepsilon -
\varepsilon') \ ,
		 						 \label{K2}
\end{equation}
where

\begin{equation}
K_d(\varepsilon, \varepsilon') = \langle \rho(\varepsilon)
\rho(\varepsilon') \rangle_{d} - \langle \rho(\varepsilon)\rangle_{d}
\langle \rho(\varepsilon') \rangle_{d} \ ,
							\label{K21}
\end{equation}
and
\begin{equation}
K_L^{\prime}(\varepsilon, \varepsilon') =
\Big\langle\langle \rho(\varepsilon) \rangle_d \langle
\rho(\varepsilon') \rangle_{d}\Big\rangle_{L} - \Big\langle \langle
\rho(\varepsilon) \rangle_d \Big\rangle_L
\Big \langle \langle \rho(\varepsilon') \rangle_d  \Big\rangle_L \ .
								\label{K22}
\end{equation}
It follows that the magnetization contains two related contributions.

The first term in eq.~(\ref{K2}) is a size-average of the
disorder-autocorrelation $K_d$ of the DOS. It is this
correlation function which is considered in the diffusive regime and which
gives the only contribution to $K(\varepsilon,
\varepsilon')$. In that case size averaging is redundant.
 The second term, $K_L^{\prime}$ , is a size-autocorrelation of the
disorder-averaged DOS:

\begin{equation}
\langle \rho(\varepsilon) \rangle_{d} = {1 \over \pi} \sum_k
{\hbar/2\tau_e \over (\varepsilon - \varepsilon_k)^2 +
(\hbar/2\tau_e)^2 } \ ,
                                                                 \label{DOS}
\end{equation}
where $\tau_e$ is the elastic collision time and  $\varepsilon_k$
are the energy levels of the clean system.
When $l_e \leq L$, the energy scale corresponding to elastic mixing of the
levels is larger than
the corresponding one-dimensional level spacing (the largest
characteristic scale for DOS modulations), and the
size-autocorrelation of $\langle \rho(\varepsilon) \rangle_{d}$
vanishes. This term, however, becomes important in the
ballistic regime. Then, the relative importance of both
autocorrelation functions $\Big\langle K_d(\varepsilon -
\varepsilon')\Big\rangle$ and $K_L^{\prime}(\varepsilon -
\varepsilon')$ depends on the value of $\varepsilon -
\varepsilon'$ and of the disorder\cite{altland}. For intermediate disorder
$l_e \gtrsim L$, both contributions to $\langle \delta N^2 \rangle$
(hence to the magnetization)
are expected to be important. In the limit of vanishing disorder (clean
system), the first term in eq.~(\ref{K2}) is suppressed and
the second term reduces to the
contribution recently addressed by Ullmo {\it et
al.}\cite{ullmo} and by von Oppen\cite{oppen2}.
 Indeed, these authors have actually
calculated the energy-autocorrelation of the DOS $\rho_0$ of the clean
system:
\begin{equation}
K_0(E) = \langle \rho_0(\varepsilon)
\rho_0(\varepsilon +E) \rangle_{\varepsilon} - \langle \rho(\varepsilon)
\rangle_{\varepsilon}^2

                                                                   \label{K3}
\end{equation}
As the expressions for the DOS and correlations thereof
contain oscillatory factors  of the form $\cos k_F L$, energy averaging
plays qualitatively the same role as size averaging.
In the semiclassical calculation
proposed by these authors, the system is assumed to be clean
(only reflection on the walls
is taken into account). Thus, eq.~(\ref{K3}) differs from eq.~(\ref{K22}) in
that the latter accounts for disorder. The contribution of the
trajectories in Refs. \cite{ullmo} and \cite{oppen2} is cut-off by a thermal
or inelastic length.

The physical situation which is considered
in our work is not the one of a
clean system but rather the one of a system in the ballistic regime
in which $l_e$ has been estimated to
be larger but of the order of $L$\cite{levy}. Moreover, one may envisage
a ballistic system where
 $l_e \lesssim L$, but with a transport mean free path $l_{tr} > L$. A
ratio of
$l_{tr} / l_e \sim 10-30$
is a common experimental situation. The implication of this fact is
indicated in the last section.

We employ a semiclassical
approach\cite{Bergmann84,Chakravarty,AIS} to evaluate the magnetic response
of ballistic conductors in the presence of disorder.
 Our semiclassical analysis as applied to the calculation of the
first contribution   $\Big\langle K_d(\varepsilon -
\varepsilon')\Big\rangle_L$ is outlined in section II.
 We find contributions to the
susceptibility
$\langle\chi (H)\rangle$ which arise  from long trajectories that are
sensitive to elastic scattering. However, due to subtle cancellations that
occur within this framework, contributions to the zero field susceptibility
do not depend on disorder. This is by no means obvious.
The zero field susceptibility is given by
\begin{equation}
\chi(H=0)=+|\chi_L|\alpha k_FL                              \label{ki2}\,.
\end{equation}
It is paramagnetic and includes an enhancement factor $\alpha k_FL$ with
respect to  the Landau susceptibility, where the numerical factor $\alpha$
is estimated below. A similar analysis
for Aharonov-Bohm geometries is carried out in section III, where we point
out major differences with respect to simply connected geometries
and discuss their origin.
 Indeed the magnetization of a ballistic cylinder threaded by an
Aharonov--Bohm flux {\em does} depend on disorder\cite{altland2}.
In the last section we discuss our result in the case of a magnetic
field in view of the experimental data.  We argue that our analysis  is in
qualitative agreement with the
experimental data of Ref.~\cite{levy}.
 We outline our predictions
concerning further measurements and discuss the range of
validity of our analysis. We comment on additional contributions to the
magnetization vis--a--vis other recent works \cite{ullmo,oppen2}.

\section{Semi-classical Analysis : THE CASE OF A MAGNETIC FIELD}

Consider a square quantum dot of size $L$. The electrons move freely
 till they are reflected elastically from a boundary or till
they are scattered elastically from an impurity. The reflexion
on the boundary is assumed to be specular, whereas scattering
from the impurities is supposed to be isotropic, the
scattering angle being uniformly and randomly distributed. (We comment on a
possible generalization in Section IV).
We consider the ballistic situation described in the introduction
where the mean free path $l_e$ between
two scattering events is much larger than the system size
$L$, but still $l_e \ll L(k_F L)$. Thus an electron is
typically reflected many times by the walls before it is
scattered by an impurity.

Our analysis is based on the calculation of the semiclassical
probability $p(t)$ of returning to the origin at a given time
$t$. The two-level correlation function
$K_d(\epsilon_1-\epsilon_2)=K_d(|\epsilon_1-\epsilon_2|)$  can then
be obtained by Fourier transforming $|t| p(t)$. This
sum rule has been obtained by Argaman {\it et al.}\cite{AIS}. Upon double
integration
 over an energy range $\Delta E$, one obtains the contribution of
$K_d$ to the number variance $\langle \delta N^2(\Delta E,H) \rangle$:

\begin{eqnarray}
\langle\delta N^2(\Delta E)\rangle&=
&\int_0^{\Delta E}\int_0^{\Delta E}K_d(\epsilon_1,\epsilon_2)\,d\epsilon_1
d\epsilon_2\nonumber\\ &=
&2\int_0^{\Delta E}(\Delta
E-\epsilon)K_d(\epsilon)\,d\epsilon\nonumber\\ &=
&4\int_0^\infty\,dt\frac{\hbar^2
p(t)}{t}(1-\cos(\frac{\Delta E
t}{\hbar}))
				     \label{dN2}\,.  \end{eqnarray}
Thus, the behavior of the field dependent spectral rigidity and of the
magnetization requires the knowledge of the field dependence of
$p(t)$.

In the present analysis we consider the contribution of
long trajectories of length ${\cal L} = v_F t \gg l_e$, so that we
calculate $p(t)$ for $t \gg \tau_e$, where $\tau_e = l_e / v_F$ is the
elastic collision time. We comment on the role of short trajectories in
Section IV.
For
such long trajectories  we assume that the end
point is uncorrelated with the starting point and that
it is uniformly distributed in the dot. Moreover, we assume that the
magnetic phase distribution (see below) is unaffected by the constraint that
only returning trajectories are to be considered.
Therefore, rather than considering the contribution of
returning trajectories of length ${\cal L}$ (in real space), one may consider
{\it all} trajectories of this length, and multiply their
contribution by a small phase space volume representing the
fraction  of those which actually return.

In the absence of
an applied field, the probability  of returning to the origin
at time $t$ is thus a constant and it is
given by (see e.~g.~ \cite{AIS})
\begin{equation} p(t)\equiv
p_0=2\frac{1}{(2\pi\hbar)^2}                      \label{p0}\,,
\end{equation}
The factor $2$ in this equation represents an
enhancement due to an interference of a trajectory with its
time reversed image (a ``Cooperon contribution'').
For a finite magnetic field, the time reversed trajectories pick
up additional phases which finally lead to the destruction of
the Cooperon contribution, as it is well known in the diffusive
regime\cite{Bergmann84,Chakravarty,AIS}. We now describe this
magnetic field effect
on the
semiclassical return probability $p(t)$, in the ballistic
regime, in the limit of long trajectories $t \gg \tau_e$.
Estimates of the field scales for which our analysis is valid
are given below. In any case, the field is sufficiently weak,
such that the bending of the trajectories can be neglected: $ H L^2 \ll
\phi_0$.
Thus the only effect of the
field consists of introducing an additional phase for
each trajectory $j$ , $\varphi_j = {2 \pi \over \phi_0} \int_j \bf
A d \bf l$, where the integral is taken along the $j$-th
trajectory.

It is convenient to describe such a
trajectory within an extended zone scheme
(Fig.~1.~A), in which boundary scattering does not occur.
It is constructed by repeatedly reflecting the dot, including
the impurity configuration, with respect  to its boundaries. According to
this picture,
the trajectory consists of straight segments of mean length
$l_e$ joining different impurities
 or images of impurities.

We next
apply a uniform perpendicular magnetic field in the
$z$--direction. Within the Landau  gauge the vector potential
in the original dot
is ${\bf A}=-H y \hat{\bf x}$. The vector potential in the
extended zone scheme is obtained by reflecting ${\bf A}$ in the
same way as the dot itself. As a result, the phase accumulated
by the electron is the same in both reduced and extended
pictures. The new vector potential is shown schematically in
Fig.~1.~B.
In this scheme, the electron is moving in a staggered magnetic
field. {\bf A} is now a periodic step function along the
$x$-direction and a periodic sawtooth function in the
$y$-direction. It may be written as sum of its Fourier components
%\begin{eqnarray} {\bf A}=-\frac{16 H L \hat{\bf
%x}}{\pi^3}\sum_{m,n=0}^\infty&&\frac{(-1)^n}{(2 m+1)(2
%n+1)^2}\sin((2 m+1)\frac{\pi x}{L})\nonumber\\ &&\sin((2
%n+1)\frac{\pi y}{L})\,.
					 %\label{A} \end{eqnarray}\\

\begin{equation} {\bf A}=-\frac{16 H L \hat{\bf
x}}{\pi^3}\sum_{m,n=0}^\infty\frac{(-1)^n}{(2 m+1)(2
n+1)^2}\sin((2 m+1)\frac{\pi x}{L})\sin((2
n+1)\frac{\pi y}{L})\,.
					 \label{A} \end{equation}\\
Consider now a trajectory of total length ${\cal L} \gg l_e$
 consisting
of numerous straight segments in this extended zone picture.
We assume that all the segments have the same length $l_e$. The
number of steps of a given trajectory of time $t$ is thus
$N = {\cal L} /l_e = v_F t/l_e = t/\tau_e$. For long
trajectories, the error made by  considering discrete times is
not substantial.

We now present an algorithm for constructing all trajectories
$j$ of length ${\cal L}$ and calculating their associated phases
$\phi_j$. We first observe that all such  trajectories come
with the same modulus of probability amplitude. This is because
we assume isotropic scattering by uncorrelated
impurities. The phase factor associated with the $j^{th}$
trajectory may be decoupled into a product of $N$ terms
corresponding to the $N$ steps.
\begin{equation}
e^{i \phi_{j}} =
e^{i \alpha_{1,j}}
e^{i \varphi_{1,j}}
e^{i \alpha_{2,j}}
e^{i \varphi_{2,j}} .....
e^{i \alpha_{N,j}}
e^{i \varphi_{N,j}} \ \ .
						\label{phase}
\end{equation}
The factors $\alpha_{i,j}$ depend on the particular impurity
configuration, and on the given path in consideration.
$\varphi_{i,j}$ are the magnetic phases accumulated at each
step $i$ of the $j^{th}$ trajectory.

In order to evaluate the associated probability, we have to
calculate $| \sum_j e^{i \phi_j} |^2$, summing over all
trajectories $j$ of a given length ${\cal L}$, i.e. of a given
time $t$. As is well known from the theory of weak
localization\cite{Bergmann84,Chakravarty}, there are only two
types of contributions that survive in this sum.
One type consists of a product of $\phi_j$ with its complex
conjugate ("diffuson contribution"). The other type ("Cooperon
contribution") consists of products of the type $e^{i
\phi_j} e^{-i
\phi_{j^{\prime}}}$, where $j$ and $j^{\prime}$ denote time
reversed trajectories. In either case the $\alpha$ phases
cancel out, and the remaining diagonal terms obtained are
insensitive to disorder averaging. The magnetic field
dependence drops out in the diffuson-like terms. For each
Cooperon-like term,
the associate phase is multiplied by a factor $2$. One gets:
\begin{equation}
\langle| \sum_j e^{i \phi_j} |^2\rangle = 2\sum_j (1 + \cos2
\varphi_j) \ \ \ \ \ \ \
\mbox{with} \ \ \ \ \ \ \varphi_j = \sum_{i=1}^N \varphi_{i,j} \ \ ,
						     \label{phase2}
\end{equation}
so that eventually $p(t)$ is now given by
\begin{equation}
p(t) = {p_0 \over 2} (1 + \langle \cos2
\varphi_j \rangle)    \ \ \ \ \ \ \
\mbox{with} \ \ \ \ \ \langle \cos2 \varphi_j \rangle =\frac{1}{M} \sum_j
\cos2 \varphi_j\,,
 \label{phase3}
\end{equation}
and $M$ is the number of trajectories of length $v_Ft$.
The phase factor $e^{2 \varphi_{i,j}}$, related
 to the $i^{th}$ step of the $j^{th}$
trajectory, is a function of the angle $\theta_{i,j}$ (Fig.~1.~A), of the
length of the step, $l_e$, and of the starting point $(x_{0(
i,j)},y_{0
(i,j)})$, equal to the end point of the previous step.
For that reason the distribution function,
$P_i(2\varphi_{i,j})$, of the phase associated with the
$i^{th}$ step, is coupled to the previous step. Nevertheless,
 we are dealing, in principle, with trajectories that consist of a large
number of steps. Since the ``memory'' of the coordinates of a given
step decay fast with the following steps, we shall make the
assumption that $P_i(2\varphi_{i,j})$ is completely decoupled
from the other steps. Technically we compute $P_i(2\varphi_{i,j})$ by
averaging
over $\theta$ and over all possible coordinates $x_0,y_0$. The average over
trajectories $\langle \cos(2 \varphi) \rangle$ is therefore given by

\begin{eqnarray}
 \langle \cos(2 \varphi) \rangle&=&{\rm Re}\left[\int_{-\infty}^\infty
P_{i}({2\varphi})e^{i2{\varphi}}\,d({2\varphi})\right]^{{\cal
L}/l_{e}}\nonumber\\
%&=&{\rm Re}\left[\frac{1}{2\pi}\int_{-\infty}^\infty\int_{-\pi}^{\pi}
%(\delta(\tilde{\phi}- \phi(\theta))e^{i2\tilde{\phi}}\,d\theta)\,d(
%\tilde{\phi})\right]^{{\cal L}/l_{e}}\nonumber\\
&=&{\rm
Re}\left[\int_{-\pi}^{\pi}\frac{d\theta}{2\pi}
e^{i2\varphi(\theta)}\right]^{{\cal L}/l_{e}}
\equiv{\rm Re}\,(\zeta(H))^{{\cal L}/l_{e}}             \label{Re}\,,
\end{eqnarray}
 $\varphi(\theta)$ being the phase
accumulated along an elementary step of length $l_e$.
The scattering angle at a given step, $\theta$, (cf.~Fig.~1.~A),
is assumed to be  uniformly distributed (a consequence of
isotropic scattering). The function $2\varphi(\theta)$
(associated with a segment of length $l_{e}$) is evaluated by
inserting eq.~(\ref{A}) into  the expression for the line
integral.

In the presence of a magnetic field, the return probability becomes, using
${\cal L}/l_e = t/\tau_e$:

\begin{equation} p(t)=\frac{1}{2}p_0 (1
+{\rm Re}\,\zeta(H)^{t/\tau_e})\label{pH}
\end{equation}

The information regarding the field dependence of $p(t)$ is contained in the
field
dependence of the factor $\zeta(H)$. In addition, inelastic scattering
which introduces an additional dephasing for long trajectories reduces the
field dependent contribution
$\Delta p(t)$. Introducing an inelastic
cutoff \cite{dupuis}, $\gamma$, we obtain
\begin{equation}
\Delta p(t)=\frac{1}{2}p_0\zeta(H)^{\frac{t}{\tau_e}}e^{-\frac{\gamma
t}{\hbar}} \label{deltap}\,.
\end{equation}
Inserting this expression into eq.~(\ref{dN2}), we find that the level
fluctuation number can be written as the sum of a field independent
and a field dependent term. The latter reads
\begin{equation}
\langle \delta \tilde N^2(\Delta E,H)\rangle_d=p_0\hbar^2\lg\frac{(\Delta
E)^2+ (\frac{\hbar}{\tau_{e}}\lg\frac{1}{\zeta}+\gamma)^2}
{(\frac{\hbar}{\tau_{e}}\ln\frac{1}{\zeta}+\gamma)^2}\label{dN2.2}
\end{equation}\\
Note that this result is formally the same as the one known for the
diffusive regime\cite{AS} in the zero mode approximation.
 Strictly
speaking eq.~(\ref{deltap}) holds
only for $t\gg\tau_{e}$. The error in substituting eq.~(\ref{deltap}) into
eq.~(\ref{dN2}) (neglecting the short time cutoff) can be shown to be small.
Insertion of eq.~(\ref{dN2.2})  into  eq.~(\ref{M}) yields the
average susceptibility $\langle\chi\rangle$

\begin{equation}
\langle\chi(H)\rangle=
{\Delta \over \gamma} { p_0 \hbar^3 \over 2 \tau_e}
{\partial^2  \ln 1/\zeta \over \partial H^2}\,.
\label{ki3}
\end{equation}

 We now proceed by calculating    $\zeta(H)$, given the vector potential
eq.~(\ref{A}) in the extended zone picture.
It turns out that $\varphi(\theta)$ is very well approximated by including
only the first harmonics of $A_x(x,y)$ ($m=n=0$ in eq.~(\ref{A})). We
checked this
numerically by calculating the low field expansion of $p(t)$ up to $m=n=2$.
The inclusion of higher harmonics did not lead to any significant modification
 of the resulting susceptibility.
Only its amplitude was changed by  an order of $1\%$ for $l_e
\sim 100L$.
This is not surprising, since the contribution of
the higher harmonics
 decay fast due to the oscillatory nature of the integrand
in  the line integral.
We therefore  restrict  ourselves in the following to the lowest harmonics,
obtaining
\begin{eqnarray}
2\varphi(\theta)&=&-2\cdot 2\pi\frac{8 H L}{\pi^3 \phi_0}\int_0^{l_e} ds
\cos\theta \\
 &&\big[\cos\left(\frac{\pi}{L}(x_0-y_0)+\frac{\pi
s}{L}(\cos\theta-\sin\theta)\right)\nonumber\\
&&-\cos\left(\frac{\pi}{L}(x_0+y_0)+\frac{\pi
s}{L}(\cos\theta+\sin\theta)\right)\big]              \label{AA}
\end{eqnarray} for a segment starting at $(x_0,y_0)$,
and ending at  $(x_0+l_{e}\cos\theta,y_0+l_{e}\sin\theta)$.
The weak field behavior and in particular the zero-field susceptibility may
be obtained by expanding $\zeta(H)$ up to second order in $H$:
\begin{equation} \zeta(H)\simeq 1-\int_{-\pi}^{\pi}
\frac{d\theta}{4\pi}(2\varphi(\theta))^2\equiv 1-c
\frac{H^2L^4}{\phi_0^2}                              \label{zeta(H)}\,.
\end{equation}
Averaging over $x_0,y_0$ we obtain
\begin{equation}
c=\frac{256}{\pi^5}\int_{0}^{l_{e}/L}ds\int_{0}^{l_{e}/L}ds'\int_{0}^{\pi}
d\theta\,\cos(\pi\sqrt{2}(s'-s)\sin\theta)              \label{c1}\,,
\end{equation}
Note that the basic
period of ${\bf A}(x,y)$ is $2L$. As the starting point  of all segments are
distributed over the
 extended zone, it is convenient to perform averaging over $x_0,y_0$
in the interval $0 \leq x_0,y_0 \leq 2L$. From eq.(\ref{c1}), one obtains:

\begin{equation}
c=\frac{512}{\pi^4}\int_0^{l_{e}/L}du\,
J_0(\sqrt{2}\pi u)(\frac{l_{e}}{L}-u)\label{c2}\,,
\end{equation} where $J_0$ is a Bessel function.
This integral can be solved exactly in terms of Struve and Bessel functions.
In the limit $l_e \gg L$, one finds:

\begin{equation}
c\simeq\frac{256\sqrt{2}}{\pi^5}\frac{l_{e}}{L} \ \ \ \ \ \ \ l_e \gg
L\label{c3} \end{equation}

Inserting this behavior in eqs. (\ref{ki3},\ref{zeta(H)})
and considering
an energy window $\Delta E\simeq \frac{\hbar}{\tau_{e}}\gg\Delta$, we obtain
\begin{eqnarray}
%% FOLLOWING LINE CANNOT BE BROKEN BEFORE 80 CHAR
\langle\chi(H=0)\rangle&=&\frac{\Delta}{\gamma}p_0\hbar^3\frac{2c}{\tau_{e}}\frac{L^4}{\phi_0^2}=\frac{1536\sqrt{2}}{\pi^{8}}\frac{\Delta}{\gamma}|\chi_L|k_FL\nonumber\\
\simeq
&&0.23\frac{\Delta}{\gamma}|\chi_L|k_FL\,,\label{ki0}
\end{eqnarray}
where  $\chi_L$ is the Landau susceptibility. For spinless electrons, it is
given by  $\chi_L=-\frac{e^2L^2}{24\pi m c^2}$, whereas for spin $1/2$
electrons, $\chi$ and $\chi_L$ have to be multiplied by $2$.
The reason why we restrict ourselves to  energy windows $\Delta
E\simeq\frac{h}{\tau_e}$ is the following. Let us consider first the
contribution of unstable trajectories, which are adequately accounted for by
our analysis. We recall that we are interested in the field dependent part
of $\langle \delta N^2\rangle$ only. The stronger the field the larger is
the energy range over which level--level correlations are affected by $H
$, i.e., the larger is the energy window $\Delta E$. The fact that $\Delta
E$ does not exceed $\frac{\hbar}{\tau_e}$ in our analysis (corresponding to
trajectories ${\cal L}>l_e$) restricts the strength of the field for which
our analysis is valid ($H<H_3$, see below). A seperate question is the
contribution of continuous, marginally stable families of trajectories
(including short ones) that have been addressed in the analyses of Refs.
\cite{ullmo} and \cite{oppen2}. We comment on these contributions in Section
IV.

The most remarkable feature in this result is that $\chi$ is {\it disorder
independent}. As we will see in the next section, this is in sharp contrast
with the behavior of the same quantity in the case of an AB flux, which does
depend on disorder\cite{altland2}. This independence on disorder is due to
to a subtle cancellation effect arising from the spatial dependence of
$A_x(x,y)$.
The main contribution to the coefficient $c$ in eq.~(\ref{c1}) comes from
segments oriented at angles $\pm \pi /4$, thus segments along which the
vector potential is either always positive or always negative. They lead to
a phase $\varphi \sim l_e/L$. However, only segments within an angle $\pm
L/l_e$ around $\theta = \pm \pi /4$ gives such a large contribution to the
integral over $\theta$, such that $c$ is proportional to $l_e/L$.
Therefore, the coefficient of $H^2$ in $\zeta(H)^{t/\tau_e} \sim (1 - c
{H^2 L^4 \over \phi_0^2})^{t/\tau_e} \sim 1 - c {t \over \tau_e} { L^4
\over \phi_0^2} H^2$ becomes independent of $l_e$.

 The range for validity of this weak field analysis
 is obtained from two conditions. First, we have to ensure that the $H^2$
expansion of $\zeta(H)$ is sufficient. Thus the $H^4$ term in the
expansion
must be small compared with the previous one. Evaluation of the $H^4$ term
leads to the condition

\begin{equation}
H < H_1 =\frac{\phi_0}{Ll_{e}} \ \ .\label{cond1}
\end{equation}
Secondly, we need that ${\hbar \over \tau_e} \ln({1 \over \zeta}) $ is
smaller
than $\gamma$. If the first condition (\ref{cond1}) is fulfilled, this
yields \begin{equation}
H < H_2 =\phi_0\sqrt{\gamma  \over v_F L^3} \ \ .\label{cond2}
\end{equation}
 The smallest of the two fields, $H_0 = min(H_1,H_2)$, makes
the cross-over to a regime where the magnetization becomes non-linear.
On the one hand, for $\gamma \tau_e < L/l_e$, the relevant field
is $H_2$ and depends therefore on the inelastic scattering rate.
If  $L_\phi \gg l_e$, we may express
$\gamma$ as $\hbar D/L_\phi^2$ which yields $H_2 = {\phi_0 \over
L^2}\sqrt{l_e L \over L_\phi^2}$. In the limit of vanishing inelastic
scattering, where $\gamma$ is of the order of $\Delta$\cite{dupuis}, we
obtain $H_2 ={\phi_0 \over
L^2}\sqrt{1 \over k_F L}$ which is independent of disorder. On the other
hand, for $\gamma \tau_e > L/l_e$, $H_1$ is the relevant field scale, which
is disorder
dependent, but independent of inelastic scattering. Thus, the way $H_0$
depends on disorder, depends critically on the inelastic scattering rate
$\hbar /\gamma$.

As a criterion for the  cross-over field we consider  the field for which
$\chi(H)$ changes  sign for the first time.
Anticipating the following results at finite field,
we show in Fig.~3 the dependence of $H_0 L^2/\phi_0$
on $\sqrt{\gamma \tau_e {L \over \l_e}}$ for several values of $l_e/L$. For
$\gamma \tau_e < L / l_e$, all curves collapse on a straight line, showing
well that in this regime the cross-over field is given by $H_2$. For larger
$ \gamma \tau_e$, $H_0$ becomes less and less dependent on $\gamma \tau_e$
 and finally reaches
 the asymptotic value given by $H_1$.

Let us now consider the finite field susceptibility. To this end we rewrite
$2 \varphi(\theta, x_0,y_0)$, eq.~(\ref{AA}), as

\begin{equation}
2\varphi(\theta,x_0,y_0)= {32 H^2 L^4\over \pi^2 \phi_0^2}\Big[C(\theta)
\sin\Big({\pi \over L}(x_0-y_0)-\alpha(\theta)\Big)+ C(-\theta)\sin\Big({\pi
\over L}(x_0+y_0)+\alpha(-\theta)\Big)\Big]
						       \label{AAnew}\,,
\end{equation}
with
\begin{equation}
C(\theta)  = {\Big|\sin\Big({\pi l_e \over \sqrt{2} L} \sin (\theta -{\pi
\over 4})\Big)\Big| \over |\sin(\theta - {\pi \over 4})|}
							  \label{new1}
\end{equation}
and
\begin{equation}
\alpha(\theta)  = \arctan\Big[{ \sin\Big({\pi \sqrt{2} l_e \over L}
\sin (\theta -{\pi \over 4})\Big) \over 1- \cos\Big({\pi \sqrt{2} l_e
\over L} \sin(\theta - {\pi \over 4})\Big)} \Big]  \ \ .
							  \label{new2}
\end{equation}

The function $\zeta(H)$ is given by the average of $\exp(2 i\varphi(\theta,
x_0,y_0)$ over $\theta$, $x_0$ and $y_0$.
We obtain:
\begin{equation}
\zeta(H)  =  {1 \over 2 \pi}
 \int_{- \infty}^{\infty} d\theta J_0\Big({32 H^2 L^4\over \pi^2
\phi_0^2}C(-\theta)\Big)   J_0\Big({32 H^2 L^4\over
\pi^2\phi_0^2}C(\theta)\Big)    \
\ . \label{zeta}
\end{equation}

An $H^2$ expansion of eq.~(\ref{zeta}) gives back eq.~(\ref{zeta(H)})
with the correct coefficient $c$.
The maximal value of $H$ for which our analysis yields the field dependence
of $\delta N^2$, namely for which $\frac{\partial^2}{\partial H^2}\langle
\delta N^2(\mu,H)\rangle\simeq\frac{\partial^2}{\partial H^2}\langle
\delta N^2(\frac{\hbar}{\tau_e},H)\rangle$ is given by $\langle
2\varphi^2(\Delta)^2\rangle\lesssim 1$. This implies:
\begin{equation}
H < H_3 =\frac{\phi_0}{\sqrt{L^3l_{e}}} \ \ .\label{cond3}
\end{equation}

Using eq.~(\ref{zeta}), we evaluated $\zeta(H)$ numerically for different
values of $l_e/L$ up to fields of the order of $H_3$, and deduced $\chi(H)$.
A typical example is shown in Fig.~2. Typically $\chi/\chi_0$
decreases with $H$ with an oscillatory behavior.
 The oscillations of $\chi(H)$ become more rapid with increasing $l_e$ and
may be suppressed completely if  either $l_e/L$ or $\gamma \tau_e$ become too
small. In that
case,  only the first change in sign of $\chi(H)$ survives. For larger
values of the field the susceptibility remains diamagnetic.

We have calculated the contribution of $K_d$ to the magnetic susceptibility
for squares of a given size, the average being taken with respect to
disorder configurations. For squares
of different sizes ( with $ \delta L > k_F$), one should average $K_d$ over
size to obtain the contribution of $\Big\langle K_d \Big\rangle_L$, the first
term of eq.~(\ref{K2}), to the magnetization and to the susceptibility.
This additional average is expected to smear the oscillatory behavior of
$\chi(H)$.

\section{Aharonov-Bohm geometries}

In close analogy with the case of a magnetic field, we now apply
a semi-classical picture to a ring threaded by an Aharonov-Bohm flux
$\phi_{AB}$.
As an example and for the sake of
simplicity, we describe the situation of a cylinder of
perimeter $L_x$ and length $L_y$ (Fig.~4). As in the case of a uniform
magnetic  field, we
adopt an extended zone scheme in which the cylinder is open and
repeated in both the $\hat{x}$ and $\hat{y}$ directions. Along the $\hat{x}$
direction there is
no reflection due to the periodic boundary condition imposed (Fig.~4). To
describe the flux $\phi_{AB}$ through the cylinder, the vector
potential is chosen as ${\bf A} = {\phi_{AB} \over L_x} \hat{ x}$. Since
it is not $y$ dependent, this expression remains unchanged in the extended
zone scheme. The  phase $\varphi$ accumulated along
a step of length $l_e$
forming an angle $\theta$ with the $x$ direction is simply
$\varphi(\theta) = 2 \pi {A l_e \over \phi_0} \cos\theta = 2 \pi
{\phi_{AB} l_e \over \phi_0 L_x}\cos\theta $.
We consider an ensemble $\{ j \}$ of trajectories of the same length $\cal
L$ made of $N ={\cal L} / l_e$ steps each of length $l_e$, with random
orientations $\theta_{i,j}$. Contrary to the uniform field case,
the returning trajectories must enclose an integer number of
the external AB flux $\phi_{AB}$ implying that the total phase $\varphi_j$
accumulated
along a trajectory $j$ is quantized $\varphi =2 \pi m \phi_{AB} /
\phi_0$. We now use the same hypotheses made in the case of a magnetic
field, i.e. long trajectories and uncorrelated steps. One  has to evaluate the
average \begin{equation}
S = \sum_m \langle \cos 2 \varphi_j \, \delta (\varphi_j - 2 \pi m
{\phi_{AB} \over \phi_0})
\rangle\,. \label{AB1}
\end{equation}
Writing the delta function as
\begin{equation}
\delta (\varphi -  \varphi_m) = {1 \over 2 \pi} \int_0^{2 \pi} e^{i
\alpha(\varphi -
\varphi_m)} d\alpha
\label{AB2}
\end{equation}
and performing the integration over each variable of the $N$ scattering
angles, $\theta_i$, one
finds, after a straightforward calculation:

\begin{equation}
S = \sum_m \int_0^{2 \pi} {d\alpha \over 2 \pi} J_0\Big( ( 2+\alpha)
{2 \pi l_e\over L}{\phi_{AB} \over \phi_0}  \Big) \exp(-2 i \pi m \alpha
{\phi_{AB}
\over \phi_0})\,. \label{AB4}
\end{equation}
Using  the identity

\begin{equation}
\sum_{m=-\infty}^\infty
\exp(-2 i \pi m  x)    = \sum_{n=-\infty}^\infty \delta(x-n)
 \label{AB5}
\end{equation}
one obtains
\begin{equation}
S  = \sum_n \Big(J_0[{2 \pi l_e \over L_x} (n+2 {\phi_{AB}
\over \phi_0})]\Big)^{{\cal L}/l_e} \,.\label{AB6}
\end{equation}
For small flux $\phi_{AB} /\phi_0 < L_x/l_e$, the  flux dependence of
$\langle\delta \tilde{N}^2(\Delta\epsilon,\phi_{AB})\rangle$ has the same form
as in eq.~(\ref{dN2.2}) with  $\zeta(\phi_{AB}) \simeq 1-4 \pi^2 ({l_e
\phi_{AB} \over
L_x \phi_0})^2$.

In the limit where $\phi_{AB} /\phi_0 < {L_x \over l_e} \sqrt{\gamma
\tau_e} < {L_x \over l_e}$ ( these inequalities correspond to $H <H_2 <
H_1$ in the case of a magnetic field), one finally obtains for the linear
expansion of the average canonical current
\begin{equation}
\langle I \rangle = - {\Delta \over 2} {\partial \over \partial
\phi_{AB}}\langle \delta \tilde{N}^2(\mu,\phi_{AB}) \rangle = {2 \over \pi}
{\Delta\over\gamma}  I_0 {l_e
\over L_x} {\phi_{AB} \over \phi_0}
\label{AB7}
\end{equation}
with $I_0 = e v_F /L_x$. Although the electronic motion is not
diffusive ($l_e > L_x$), the persistent current
formally assumes the same expression as in the diffusive regime \cite{AGIM}.
For $H>H_2$, the current drops as $1/\phi_{AB}$, as it has been found in the
diffusive regime \cite{AGIM}. This result is
valid throughout the entire ballistic regime and can even be extrapolated up
to the crossover to the clean regime, where $l_e \sim k_F L_x
L_y$. In that case, one finds a saturation of the small flux current to
$\langle I \rangle \sim I_0 k_F L_y \phi_{AB} / \phi_0$.
 The average susceptibility can be rewritten as $E_c
/ \phi_0$ where $E_c$, although still being formally given
by $E_c \sim {\hbar v_F l_e \over L_x^2}$, is {\it not} anymore the
inverse of the traversal time   $\tau_{fl} = L/ v_F$.
It is actually a
non-trivial result that the energy scale $E_c$ still drives the
small flux sensitivity of the energy levels, even in the
ballistic regime\cite{altland2}.

\section{Conclusion}

To summarize, the central point of our analysis was based on the observation
that when the condition $l_\phi\gg l_{e}$ is met,
  the weak field behaviour carries the signature of the {\em
long}, disorder dependent trajectories.
In the case of a magnetic field, we find an average zero field susceptibility
$\langle |\chi| \rangle$
which is disorder independent and proportional to ${\Delta \over \gamma}
\chi_L k_F L$ up to a numerical factor of order one. In the case of an
Aharonov-Bohm flux, the susceptibility is disorder dependent and
proportional to the mean free path as in the diffusive regime. Significant
deviations are found for stronger fields, where the non--linear
susceptibility may change sign and is, in general, disorder dependent.

It is interesting to compare our result  for the case of a uniform magnetic
field with recent experiments.
We first note that in the presence of inelastic scattering ( or at finite
temperature), the cut-off $\gamma$ may be written as $\hbar /
\tau_{\varphi}$, where
$\tau_{\varphi}$ is a typical inelastic scattering (or dephasing) time. For
long
trajectories, if $l_{\varphi} \gg l_e$, one has $l_{\varphi}^2 = D
\tau_{\varphi}$ so that $\langle\chi\rangle/ |\chi_L| \sim l_{\varphi}^2 /
L l_e$. The
magnetic susceptibility of two dimensional ballistic dots (squares)
has recently been measured  by L\'evy et al.~\cite{levy}. They have obtained
$\frac{\langle\chi(H=0)\rangle}{|\chi_L|}\simeq  k_FL$. In their experiments
$l_{e} \simeq L-2 L$, $l_\phi\simeq 3L-10L$.
With these parameters eq.~(\ref{ki0}), pushed to the limit of its validity,
yields ($l_{e}=1.5 L$, $l_\phi=8L$) $\langle\chi\rangle\simeq 250
|\chi_L|$, in rough agreement with the experiment. Evidently, a more
detailed comparison with our theory (e.~g., measurements of the
crossover field from linear susceptibility, which is sensitive to disorder)
is needed before
the validity of our picture can be confirmed.

We have calculated here the contribution to the magnetization of the  size
averaged  disorder-autocorrelation function $\Big\langle K_d
\Big\rangle_L$. To this contribution, one should add the contribution of
the
size-autocorrelation of the disorder-averaged DOS $K^{\prime}_L$. This
latter quantity has
been investigated in Refs. \cite{ullmo,oppen2} {\it in the limit of no
elastic  disorder} (but with the inclusion of an inelastic or thermal cut-off).
Following eq.~(\ref{K2}), we stress that  these two contributions exist and
contribute to the DOS-autocorrelation function and therefore to $\langle
\delta \tilde{N}^2\rangle$ and its derivatives.
One may thus conclude that the contribution of Refs.~\cite{ullmo} and
\cite{oppen2} (corrected for disorder) should
be added to that found in the present work  to yield the average
susceptibility. Indeed, these two contributions have very different disorder
and field dependence.  The experimental situation probably
corresponds to a case close to the borderline between ballistic and
diffusive regimes, where the two
contributions turn out to be roughly of the same order of magnitude. We note,
though, that the contribution arising from $K_L'$ (namely, the one related
to Refs.~\cite{ullmo} and \cite{oppen2}) may be, after all, negligible in
the experimental situation addressed in Ref.~\cite{levy}: In the presence of
elastic scattering, i.e. at finite $l_e$, the Green function
averaged on disorder decreases exponentially as $\exp(-L/l_e)$, and so does
the associated autocorrelation function $K_L'$. This is an important point,
because the experimental situation
 corresponds, most probably, to the
case where $l_e$ (corresponding to a small angle scattering) is shorter
than the transport mean free path, $l_{tr}$, namely $l_{tr}\gtrsim L> l_e$.
Our contribution is significant as long as $l_{tr}\gtrsim L$ (the above
analysis assumes $l_{tr}=l_e$, but if they differ it is $l_{tr}$ which
counts).

{\it Acknowledgments: }
We have benefitted from useful discussions with A.~Alt\-land, N.~Argaman,
O.~Bohigas,
R.~A.~Jalabert, K.~Richter, and D.~Ullmo. In particular we are grateful to
D.~Mukamel for pointing out to us the usefulness of the expansion
eq.~(\ref{A})  in the present context.
Y.~G.~acknowledges  the
hospitality of H.~Bouchiat and G.~Montambaux in Orsay. This research
 was supported in part by the Claussen
Stiftung, the German--Israel Foundation (GIF), the
U.~S.~--Israel Binational Science Foundation (BSF), and the EC Science
program no. SCC--CT90--0020.\\

\begin{figure}
Fig.~1.~A.
A multiply reflected semiclassical trajectory (thick line), and its image
within the extended zone scheme. A dot denotes the location of a scatterer.
\label{ext}
\end{figure}
\begin{figure}

FIG.~1.~B. Vector potential within the extended  zone scheme (arrow
lengths
are proportional to $|A_x|$). Alternating directions of the staggered field
are indicated.
\end{figure}
\begin{figure}
FIG.~2. The field dependent susceptibility calculated numerically
(Eqs.~(\ref{M}), (\ref{dN2.2}) and (\ref{zeta})) for
$H<\frac{\phi_0}{\sqrt{l_{e} L^3}}$. Here $l_{e}/L=50.3$. Note that
$\langle\chi(H)\rangle$ changes sign for $H>\frac{\phi_0}{Ll_{e}}$. These
oscillations will be partially smeared out due to fluctuations in the
value of $l_{e}$.
\end{figure}

\begin{figure}
FIG.~3 The cross-over field $H_0$ from linear to non--linear susceptibility,
evaluated as the field for which $\chi(H)$ changes sign for the first time,
as a function of $\sqrt{\gamma\tau_e\frac{L}{l_e}}$. Full circles correspond
to $l_e/L=50.3$, stars to $l_e/L=20.2$ and crosses to $l_e/L=10.3$.
\end{figure}

\begin{figure}
FIG.~4 Aharonov Bohm cylinder (extended zone)
\end{figure}

\end{document}